\documentclass[rnaas]{aastex631}
\usepackage{natbib}
\usepackage{graphicx}

\accepted{September 8, 2022}
\submitjournal{RNAAS}

\begin{document}

\title{The Exoplanet Modeling and Analysis Center at NASA Goddard}
\shorttitle{Exoplanet Modeling and Analysis Center}

\author[0000-0002-8619-8542]{Joe P. Renaud}
\affiliation{University of Maryland, College Park, MD 20742}
\affiliation{NASA Goddard Space Flight Center, 8800 Greenbelt Road Greenbelt, MD 20771, USA}
\affiliation{Center for Research and Exploration in Space Science and Technology, NASA Goddard Space Flight Center, Greenbelt, MD 20771}

\correspondingauthor{Joe P. Renaud}
\email{joseph.p.renaud@nasa.gov}

\author{Eric Lopez}
\affiliation{NASA Goddard Space Flight Center, 8800 Greenbelt Road Greenbelt, MD 20771, USA}

% Science post-baccs/alum in alpha Order

\author[0000-0002-2072-6541]{Jonathan Brande}
\affiliation{Department of Physics and Astronomy, University of Kansas, 1082 Malott, 1251 Wescoe Hall Dr., Lawrence, KS 66045, USA}

\author[0000-0002-7454-281X]{Carlos E. Cruz-Arce}
\affiliation{Southeastern Universities Research Association, 1201 New York Ave., NW, Suite 430, Washington, DC 20005}
\affiliation{Center for Research and Exploration in Space Science and Technology, NASA Goddard Space Flight Center, Greenbelt, MD 20771}
\affiliation{NASA Goddard Space Flight Center, 8800 Greenbelt Road Greenbelt, MD 20771, USA}

\author{Cameron Kelahan}
\affiliation{Southeastern Universities Research Association, 1201 New York Ave., NW, Suite 430, Washington, DC 20005}
\affiliation{Center for Research and Exploration in Space Science and Technology, NASA Goddard Space Flight Center, Greenbelt, MD 20771}
\affiliation{NASA Goddard Space Flight Center, 8800 Greenbelt Road Greenbelt, MD 20771, USA}

\author{Nicholas Susemiehl}
\affiliation{Southeastern Universities Research Association, 1201 New York Ave., NW, Suite 430, Washington, DC 20005}
\affiliation{Center for Research and Exploration in Space Science and Technology, NASA Goddard Space Flight Center, Greenbelt, MD 20771}
\affiliation{NASA Goddard Space Flight Center, 8800 Greenbelt Road Greenbelt, MD 20771, USA}

% Dev team in alpha order
\author{Dylan Cristy}
\affiliation{NASA Goddard Space Flight Center, 8800 Greenbelt Road Greenbelt, MD 20771, USA}

\author{Carl Hostetter}
\affiliation{NASA Goddard Space Flight Center, 8800 Greenbelt Road Greenbelt, MD 20771, USA}

\author[0000-0001-7912-6519]{Michael Dane Moore}
\affiliation{Business Integra Inc., 6550 Rock Spring Dr., Suite 600, Bethesda, MD 20817, USA}
\affiliation{NASA Goddard Space Flight Center, 8800 Greenbelt Road Greenbelt, MD 20771, USA}

\author{Apexa Patel}
\affiliation{KBR Inc., 7701 Greenbelt Road, Suite 400. Greenbelt, MD 20770, USA}
\affiliation{NASA Goddard Space Flight Center, 8800 Greenbelt Road Greenbelt, MD 20771, USA}

% Project lead
\author[0000-0002-8119-3355]{Avi M. Mandell}
\affiliation{NASA Goddard Space Flight Center, 8800 Greenbelt Road Greenbelt, MD 20771, USA}

\section{Abstract}
The Exoplanet Modeling and Analysis Center (EMAC) at NASA Goddard Space Flight Center is a web-based catalog, repository, and integration platform for modeling and analysis resources focused on the study of exoplanet characteristics and environments. EMAC hosts user-submitted resources ranging in category from planetary interior models to data visualization tools. Other features of EMAC include integrated web tools developed by the EMAC team in collaboration with the tools' original author(s) and video demonstrations of a growing number of hosted tools. EMAC aims to be a comprehensive repository for researchers to access a variety of exoplanet resources that can assist them in their work, and currently hosts a growing number of code bases, models, and tools. EMAC is a key project of the NASA GSFC Sellers Exoplanet Environments Collaboration (SEEC) and can be accessed at \url{https://emac.gsfc.nasa.gov}.

\section{Introduction}
It has been three decades since the discovery of the first extrasolar planets. In that time, the research output and publications associated with exoplanet observations, data analysis, and modeling has risen exponentially; at the same time, transformations in the technical and cultural aspects of information dissemination have made the sharing of resources, software, and model inputs much simpler and more ubiquitous. However, there has not been a comprehensive platform for cataloging, sharing, and comparing these resources in an exoplanet research context. \par

The \href{https://seec.gsfc.nasa.gov/}{Sellers Exoplanet Environments Collaboration} (SEEC) at NASA Goddard Space Flight Center has initiated the \href{https://emac.gsfc.nasa.gov}{Exoplanet Modeling and Analysis Center} (EMAC) to serve this purpose. EMAC is a web and mobile-accessible system that serves as a catalog, repository, and integration platform for modeling and analysis resources focused on the study of exoplanet characteristics and environments. At the time of writing, EMAC has cataloged over 170 tools and resources split into 11 scientific categories and numerous subcategories. In this report, we describe the design and future plans for this project. We encourage developers in the exoplanet community to submit their resources to the platform and utilize EMAC's search tools and subscription services when starting their next project. \par

\section{Exoplanet Resource Database}
\begin{figure}
\centering
\includegraphics[width=0.95\textwidth]{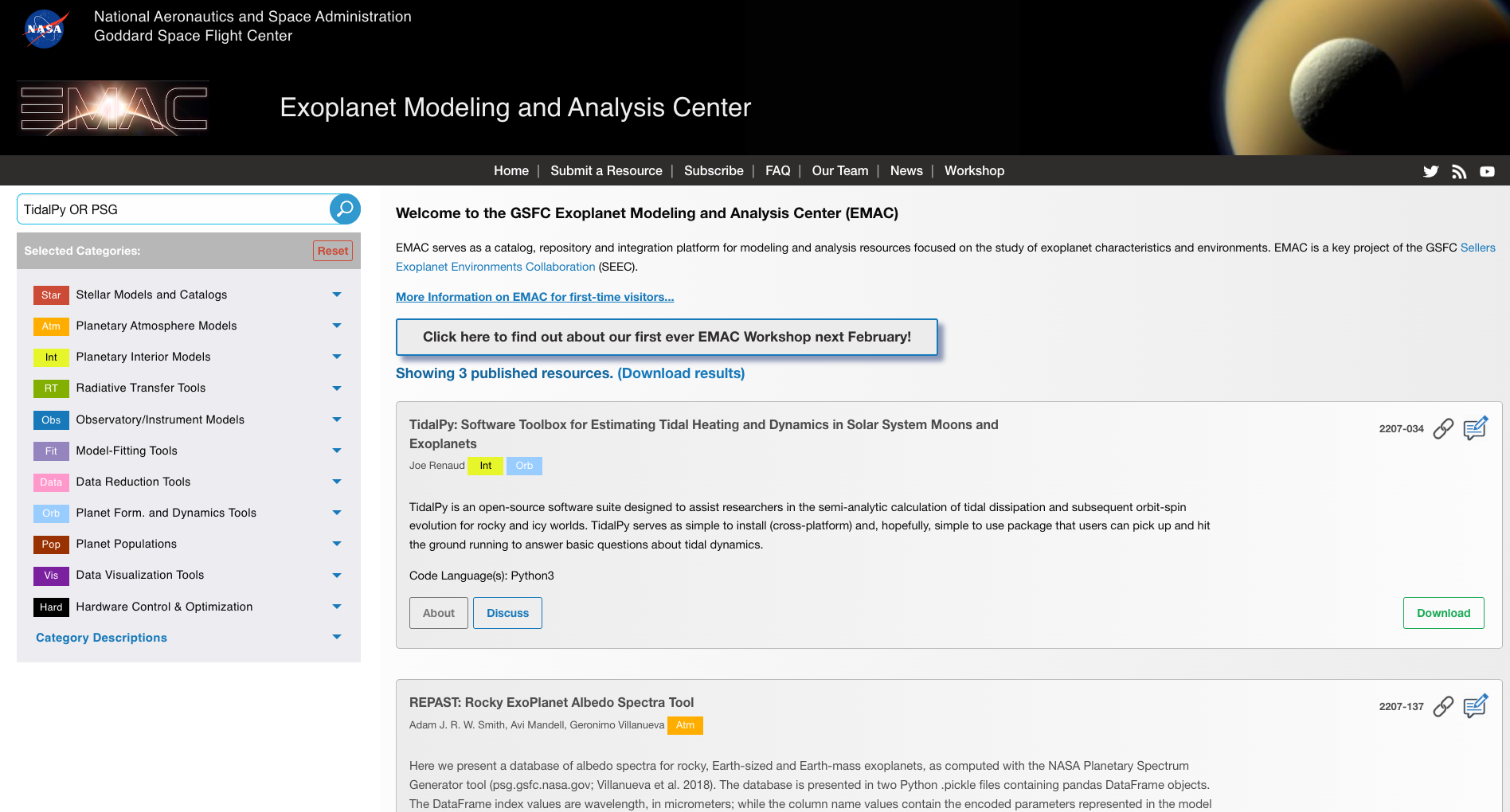}
\caption{\label{fig1} The EMAC homepage features all user-submitted resources that have been approved by the EMAC team. Major scientific themes can be seen to the left of the image. These categories can be clicked by a user to filter the resource list based on one or more topics and capabilities. A search bar is available for users to quickly find a specific tool. Each resource block shows a summary of the resource, links to applicable code repositories, documentation, and tutorials.}
\end{figure}

An EMAC-listed resource is any software, data visualization tool, or collection of model inputs or outputs related to exoplanet science. Users can \href{ https://emac.gsfc.nasa.gov/submissions/}{submit} a resource to EMAC where it will be reviewed by our team to ensure its applicability. Once approved, it will be featured on the EMAC homepage (See Figure \ref{fig1}) and will be searchable by category or keyword. Each resource is displayed to users in discrete ``resource blocks.'' These blocks contain metadata such as authors, summaries, as well as links to 3rd-party code repositories, tutorials, notebooks, and documentation. EMAC automatically checks all resources for broken links every month and works with the resource authors to update or fix these issues. \par

Every resource is provided a unique ID that follows a format similar to the ArXiv manuscript repository: "[2-Digit Year][2-Digit Month]-[3-Digit Sequential Number]". A resource can be accessed on the EMAC website utilizing its ID via the URL scheme: "https://emac.gsfc.nasa.gov?cid=UniqueID". The URL provides a means to quickly and easily share multiple types of information (code, databases, tutorials, documentation, etc.), all of which may be hosted across multiple sites and servers, with a single, permanent link. \par

The primary users of EMAC are exoplanet scientists that are looking to solve a problem but may not know the best software or dataset to use. To assist these users, EMAC offers a multi-category search function so that visitors can create filters to list only those resources that apply to their questions. Categories include ``Planetary Atmosphere Models'', ``Planetary Interior Models'', ``Planet Formation and Dynamics Tools'', ``Observatory/Instrument Models'', and many more. Each of these categories can have several sub-categories to allow for more refined searches. EMAC is always looking to expand into new areas of research, and we encourage authors who have a resource that may not fit a current category to either submit and suggest a new category or contact the team to offer new category suggestions. In addition to filtering, resources with significant heritage in common can be linked to one another using the ``Related'' button. This allows users to quickly find tools with a common code base or compare and contrast various exoplanet models with one another. \par

\section{Promoting and Supporting Resources \& Developers}
Resource development is time-intensive and scientists who publish their tools online often do not have the time or expertise to promote them beyond brief mentions in talks and papers. EMAC provides services to promote resources with little-to-no extra work on the author's end. We offer a \href{https://emac.gsfc.nasa.gov/subscriptions/}{subscription service} and a \href{https://emac.gsfc.nasa.gov/news/rss/}{RSS feed} that interested users can sign up to based on their research interests to learn when a new resource is posted, or updates are made to an existing one. Our \href{https://twitter.com/ExoplanetModels}{Twitter} account shares information about resources and connects users with developers should questions arise. The EMAC team has also created video tutorials for various exoplanet software which are then hosted on our \href{https://www.youtube.com/channel/UCLRJnT1l6CGC8aU2o2MofXw}{YouTube} channel, which has received thousands of views. \par

Based on user feedback, we have found that the most accessible tools are those that have web-based applications or graphical user interfaces. However, these can be time-consuming for authors to develop when their focus is on conducting scientific research. For that reason, EMAC offers users assistance in building out these resources, particularly for applications that utilize \href{https://jupyter.org/}{Jupyter notebooks} that can be easily opened on the web-based \href{https://mybinder.org/}{MyBinder} platform. Our team can assist researchers in putting together these web-based tutorials that showcase their software and its use cases.

\section{Future Plans}
The primary missions of EMAC are to improve the accessibility of exoplanet resources, foster inter-code comparison and validation, and provide a venue to support developers of publicly available exoplanet software resources. To further these goals, we will offer the first \href{https://emac.gsfc.nasa.gov/workshop/}{EMAC workshop} in early 2023, which will bring our community of developers and users together to present and share codes and expertise with the wider exoplanet community and highlight the exciting work being done with these software packages. We hope that these regular events, followed up by an online discussion platform, will lead to long-term networking across exoplanet disciplines and will encourage future model comparisons and new code development. \par

Going forward, we plan to host more web-based tools and visualization resources to showcase the capabilities of EMAC-listed resources and inform when one resource may be more applicable to a specific problem compared to another. While EMAC's primary audience is exoplanet scientists, we also plan to provide outreach to non-scientists and students by working with organizations to utilize a subset of EMAC-listed tools in high school and college classrooms. This learning-by-doing approach will both prepare students for a research career and educate them about the fundamentals of exoplanet science and scientific computing in general. \par

We invite the exoplanet community to \href{https://emac.gsfc.nasa.gov/submissions/}{submit} their resources, join EMAC's \href{https://emac.gsfc.nasa.gov/subscriptions/}{subscription service}, and \href{mailto:gsfc-emac@mail.nasa.gov}{connect} with our team to explore future ideas and collaborations.

\section{Acknowledgments}
EMAC is a key project of the NASA Goddard Space Flight Center's \href{https://seec.gsfc.nasa.gov/}{Sellers Exoplanet Environments Collaboration (SEEC)}, which is funded through the Internal Scientist Funding Model (ISFM) by NASA's Planetary Science Division (PSD), the Astrophysics Division (APD), and the Heliophysics Division (HPD). J.P.R, C.A.C, C.K., and N.S. are supported by NASA under award number 80GSFC21M0002.

%V\bibliography{bibliography}

\end{document}